\documentclass[seceq]{ptptex}
\usepackage{graphicx} 
\usepackage{latexsym}
\notypesetlogo
\begin{document}
  \begin{flushright}
   RUP-14-19 \\
\end{flushright}
\vspace{10mm} 
\begin{center}
\Large{ \bf TMD Parton Distributions based on Three-Body Decay Functions in NLL Order of QCD    }

\vspace{15mm}

\large{Hidekazu {\sc Tanaka} \\

 Department of Physics, Rikkyo University, Tokyo 171-8501, Japan\\

     }


\vspace{25mm}

  {\Large ABSTRACT}
  \end{center}
        
\vspace{10mm}
\def\proj{{\bf P}}  
\def\slsh#1{{#1}{\kern-6pt}/{\kern1pt}}  



        

Three-body decay functions in space-like parton branches are implemented to evaluate transverse-momentum-dependent (TMD) parton distribution functions in the next-to-leading logarithmic (NLL) order of quantum chromodynamics (QCD).  Interference contributions due to the next-to-leading order contribution are taken into account for the evaluation of the transverse momenta in initial state parton radiations. 
 Some properties of the decay functions are also examined.

 As an example, we compare our results with an algorithm proposed in Ref.\citen{rf:1}, in which a transverse momentum distributions are evaluated at the last step of parton evolutions.

  \newpage

\section{Introduction}

Parton distribution functions are important ingredients for evaluation of hadron scattering processes.    In most of calculations for the hadronic cross sections, scale dependent parton distributions for longitudinal momentum fractions are implemented, in which  the transverse momentum of partons are integrated over. 
 However, transverse-momentum-dependent (TMD) parton distributions  are more appropriate for evaluation of the transverse momentum distributions of produced particles in hadron scattering processes. 

 In order to evaluate the TMD parton distributions, one of the methods has been proposed in Ref. \citen{rf:1}, in which the transverse momentum distributions are given in the last step of parton evolutions. In this method, angular ordering conditions due to interference effects are imposed. 
This method has been extended to include the next-to-leading order terms.

 Alternatively, one can use parton shower models in order to evaluate the transverse momentum distributions as well as the scaling violation of the parton distributions.  One such algorithm has been proposed in Ref. \citen{rf:2}, in which the parton showers are generated at the next-to-leading-logarithmic (NLL) order of quantum chromodynmics (QCD) using an algorithm consisting of a model based on longitudinal momentum conservation of partons.  In this model, the scaling violation of the parton distributions is generated using only information from the splitting functions of the parton branching vertices and input distributions at a given energy.  It has been found that the method reproduces the scaling violation of the parton distributions up to their normalizations at the NLL order of QCD.

  The transverse momentum of partons are generated according to two-body decays including information of three-body decay functions for parton branching processes, which have been calculated at the NLL order of QCD.\cite{rf:3}
 The higher order terms of parton branching vertices extracted by collinear factorization naturally include interference contributions of branching processes.    It has been shown that double logarithmic terms in the three-body decay functions can be included in kinematical bound for effective two-body branching vertices, which define kinematical constraints of the two-body branching processes. In soft gluon region, the three-body decay functions reproduce the effects of angular ordering conditions. \cite{rf:3}\cite{rf:4}

In this paper, we study TMD parton distributions calculated by the algorithm implemented in Ref.\citen{rf:2} at the NLL order of QCD and the results are compared with obtained by the method proposed in Ref. \citen{rf:1}.

In section 2, we present relations between the three-body decay functions and the angular ordering condition implemented in Ref. \citen{rf:1}. Some numerical results are shown in section 3. Section 4 is devoted to summary and some comments. 

\section{Three-body decay functions and angular ordering}

Though various processes contribute to the initial state parton radiations, the important contributions are $q(S) \rightarrow q(S)+X$ and $g(S) \rightarrow g(S)+X$, where $S$ represents a parton with space-like virtuality.  Particularly, two-gluon radiation, such as
\begin{eqnarray}
 a(p,S) \rightarrow g(k_1)+g(k_2)+a(k_3,S),
\end{eqnarray}
becomes large in the soft gluon region. Here, $a$ represents a quark ($a=q$) or a gluon ($a=g$) with space-like virtuality.  
The momenta of these partons are denoted by $p$, $k_1,k_2$ and
$k_3$, respectively.  

 In this section, we consider a relation between the three-body branching for soft gluon radiation and angular ordering condition implemented in Ref.\citen{rf:1}.

  Three-body decay functions are   coefficients of the $1/(-s)$ contribution (collinear contribution) for the branching vertex at $O(\alpha_s^2)$ of QCD, which is defined by
\begin{eqnarray}
V^{(3)}_{a}=\Bigl({\alpha_s\over 2\pi}\Bigr)^2 
\delta(1-z_1-z_2-z_3)dz_1dz_2dz_3 \sum^D_{j=A}J^{[j]}_a {d(-s) \over -s},
\end{eqnarray}
where $\alpha_s$ denotes the strong coupling constant of QCD.
 
In the calculation of the three-body decay functions, the parton momenta are set as $p^2=k_1^2=k_2^2=0$ and $k_3^2=s<0$, because the collinear contributions for $-p^2, k^2_1, k^2_2 \ll -s$ are extracted. Here, $j=A - D$ indicate the types of squared matrix elements defined by the 
 structures of the propagators.\footnote{The diagrams which contribute the three-body decays are classified into following types 
according to its structure of the denominators of the squared matrix elements.\cite{rf:3}  Here we define invariants as $s_{ij}=(k_i+k_j)^2$ for $i\ne j$.    
 It should be noted that $s_{12}>0$ and $s_{13},s_{23}<0$.

\begin{description}
\item{Type [A]:}  Two same time-like propagators($M_A \propto 1/s_{12}^2$).
\item{Type [B]:}  Two same space-like propagators($M_{B1} \propto 1/s_{23}^2$ or $M_{B2} \propto 1/s_{13}^2$).
\item{Type [C]:}  One time-like propagator and  a space-like one
 ($M_{C1} \propto 1/(s_{12}s_{23})$ or $M_{C2} \propto 1/(s_{12}s_{13})$).
\item{Type [D]:}  Two different space-like propagators($M_D \propto 1/(s_{13}s_{23})$).
 \end{description}
Here, an amplitude $T_{4g}$ for the four gluon interaction is written by $ T_{4g}s_{12}/s_{12}$, thus this contribution can be included in one of four types of amplitudes. }

 The momentum fraction is defined by
\begin{eqnarray}
z_i={k_i n \over pn}, 
\end{eqnarray}
where $n$ is a light-like vector that specifies the light-cone gauge. Here, $z_1+z_2+z_3=1$ is satisfied. 

 The quantity $J^{[j]}_a$ in Eq. $(2\cdot 2)$ is written as
\begin{eqnarray}
J^{[j]}_a =  \int^{(-s)}_{M_{[j]}^2}L^{[j]}_a{dK^2_j \over K^2_j} +  L^{[j]}_a{\rm log}W^{[j]} + N^{[j]}_a 
\end{eqnarray}
for $j=A,B1,B2$ and 
\begin{eqnarray}
J^{[j]}_a =  L^{[j]}_a{\rm log}W^{[j]} + N^{[j]}_a 
\end{eqnarray}
for $j=C1,C2,D$. 
    Furthermore, $K_{A}^2=s_{12},K_{B1}^2=-s_{23}$ and $K^2_{B2}=-s_{13}$, respectively.

 In Eq. $(2\cdot 4)$, $M_{[j]}$ is a minimum mass scale of the phase space integration. Here, we define $M_{[A]}^2=l_0^2$ and $M_{[B1]}^2=M_{[B2]}^2=M_0^2$. 
  The explicit expressions of $L,N$ and $W$ in the light-cone gauge are presented in Ref. \citen{rf:3}.
 As shown there, the functions $L^{[j]}_a$ for $j=A,B1,B2$ are  the convolutions of the splitting functions of the two-body branching vertices at the LL order of QCD. The first term of Eq. $(2\cdot 4)$ is regarded as the $O(\alpha_s^2)$ term  of the LL order contribution, which should be subtracted form $V^{(3)}_{a}$ in Eq. $(2\cdot 2)$ and these are included in two-body branching vertex.   The interference terms (types [C] and [D])  are free from the mass singularity for fixed $s$. Therefore a ${\rm log}(-s/M_{[j]}^2)$ term does not appear in Eq. $(2\cdot 5)$.

As shown in Ref.\citen{rf:3}, for $z_1 \ll z_2,z_3$, the interference term gives a large logarithmic contribution as $J^{[C1]}_a \sim O(z_1^{-1}\log z_1)$.

 As a result of the subtraction of the LL order terms from  $V^{(3)}_{a}$, there is some freedom in defining the NLL order terms. 
 The three-body decay functions are  modified as
\begin{eqnarray}
J_M^{(a)}=J_0^{(a)}-J_S^{(a)} 
\end{eqnarray}
with 
\begin{eqnarray}
J_0^{(a)}=\sum_{j=A}^DJ_a^{[j]}
\end{eqnarray}
and
\begin{eqnarray}
J_S^{(a)}=\int_{l_0^2}^{(-s)f^{[A]}_a}L^{[A]}_a{dK_{A}^2\over K_{A}^2}+\int_{M_0^2}^{(-s)f^{[B1]}_a}L^{[B1]}_a{dK_{B1}^2 \over K_{B1}^2}+\int_{M_0^2}^{(-s)f^{[B2]}_a}L^{[B2]}_a{dK_{B2}^2\over K_{B2}^2},
\end{eqnarray}
 where $f^{[B1]}_a,f^{[B2]}_a$ and $f^{[A]}_a$ are the functions that depend on $z_i$. The subtracted  contributions presented in Eq.$(2\cdot8)$ are included in the kinematical constraints of the two-body branching vertices as $K^2_A<(-s)f^{[A]}_a, K^2_{B1}<(-s)f^{[B1]}_a$ and  $K^2_{B2}<(-s)f^{[B2]}_a$, respectively.
  Furthermore, one can stipulate that all the NLL contributions are absorbed in the phase space restriction for the space-like parton branch as $J_M^{(a)}=0$, which  gives a relation 
\begin{eqnarray}
{\tilde J}_0^{(a)}={\tilde J}_S^{(a)}=  L^{[A]}_a{\rm log}f_a^{[A]}+L^{[B1]}_a{\rm log}f_a^{[B1]} + L^{[B2]}_a{\rm log}f_a^{[B2]}.
\end{eqnarray}
with
\begin{eqnarray}
{\tilde J}_I^{(a)}=J_I^{(a)}-L^{[A]}_a{\rm log}(-s/l_0^2)-(L^{[B1]}_a+L^{[B2]}_a){\rm log}(-s/M_0^2)
\end{eqnarray}
for $I=0$ and $I=S$.

Therefore, the kinematical boundary for out-going virtual gluons are given as 
\begin{eqnarray}
 f^{[A]}_a={\rm exp}\left[{{\tilde J}_0^{(a)} - L^{[B1]}_a{\rm log}f_a^{[B1]} - L^{[B2]}_a{\rm log}f_a^{[B2]} \over  L^{[A]}_a}\right].
\end{eqnarray}

In this paper, we chose 
\begin{eqnarray}
 f^{[B1]}_a={z_1 \over y_3y_1},~~ f^{[B2]}_a={z_2 \over y_3y_2}
\end{eqnarray}
with $y_i=1-z_i$. As shown in Appendix A, this choice may correspond to the angular  ordering condition implemented in Ref.\citen{rf:1}.

The $z_1$ dependence of $f^{[A]}_a$ for $a=q,g$  are presented in Figs.1 and 2.  

As shown in the figuars,  the contributions of $N_a\equiv\sum_{j=A}^DN_a^{[j]}$ are small compared with the logarithmic terms, $V_a\equiv\sum_{j=A}^{B2}L_a^{[j]}\log W^{[j]}$, which are represented by dotted curves.

\begin{figure}
\centerline{\includegraphics[width=10cm]{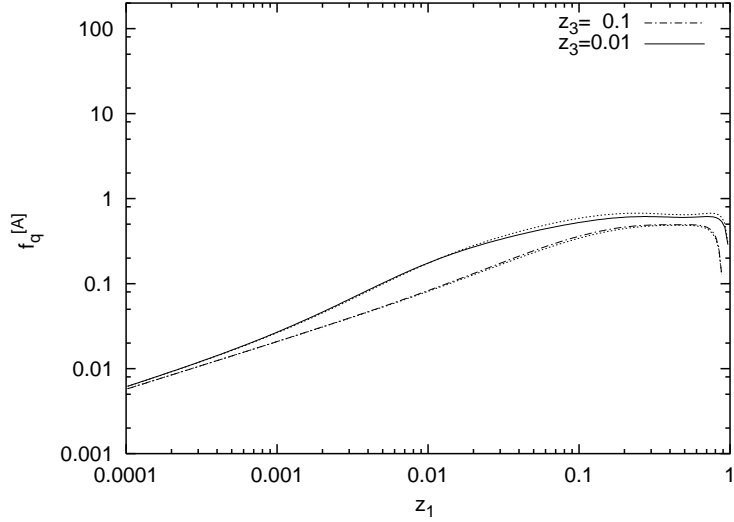}}
\caption{The $z_1$ dependence of the function $f_q^{[A]}$ with $z_3=10^{-1}$ and $10^{-2}$ are represented by the dash-dotted curve and the solid curve, respectively.  The dotted curves  represent the results without $N_q$ term.
}
\end{figure}
\begin{figure}
\centerline{\includegraphics[width=10cm]{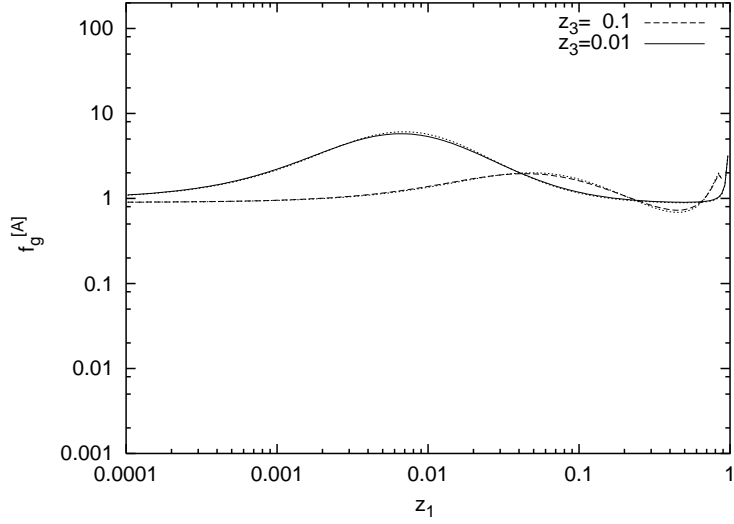}}
\caption{The $z_1$ dependence of the function $f_g^{[A]}$ with $z_3=10^{-1}$ and $10^{-2}$ are represented by the dash-dotted curve and the solid curve, respectively.  The dotted curves  represent the results without $N_g$ term.
}
\end{figure}

The $z_1$ dependence of $f^{[A]}_a$ may be understood by following consideration.

Since $N_a \ll V_a$ is numerically satisfied, we neglect the $N_a$ terms in the following consideration.

  Here,  we define
\begin{eqnarray}
K(z_i)={(1-z_iy_i)^2 \over z_iy_i},~~~P(z_i)={1+z_i^2 \over y_i}.
\end{eqnarray}
  For $z_1 \ll z_2,z_3$ gives $y_1\simeq 1,y_2\simeq z_3$ and $y_3\simeq z_2$. The arguments of the logarithmic factor in $J_a^{[i]}$ are approximated as 
 \begin{eqnarray*}
  W^{[A]} &=& {y_3 \over z_3}  ~,~~ W^{[B1]} = {y_1 \over z_3} \simeq {1 \over z_3} ~,~~
  W^{[B2]} = {y_2 \over z_3} \simeq 1~, 
\end{eqnarray*}
\begin{eqnarray}
  W^{[C1]} &=& {y_1y_3 \over z_1z_3} \simeq {y_3 \over z_1z_3}~,~~ W^{[C2]} = {y_2y_3 \over z_2z_3} \simeq 1~,~~
  W^{[D]} = {y_1y_2 \over z_3} \simeq 1~.  
 \end{eqnarray}
Therefore, types [A],[B1] and [C1] contribute to the logarithmic terms.

For  $z_1 \ll z_2,z_3$, we have
 \begin{eqnarray}
 L_q^{[A]} \simeq  -L_q^{[C1]} \simeq 2C_FC_A{P(z_3) \over z_1},~~ L_q^{[B1]} \simeq  2C_F^2{P(z_3) \over z_1}. 
 \end{eqnarray}
Here, $C_A=3,C_F=4/3$ are the color factors. Therefore,  ${\tilde J}^{(q)}_0$  for  $z_1 \ll z_2,z_3$ is approximated as\begin{eqnarray}
{\tilde J}^{(q)}_0 \simeq 2C_FC_A{P(z_3) \over z_1}\log z_1+2C_F^2{P(z_3) \over z_1}\log {1 \over z_3}
\end{eqnarray}
and
\begin{eqnarray}
{\tilde J}^{(q)}_0 - \sum_{j=B1,B2} L^{[j]}_q{\rm log}f_q^{[j]} \simeq 2C_F\left[(C_A-C_F)\log z_1+C_F\log y_3 \right]{P(z_3) \over z_1},
\end{eqnarray}
which gives
\begin{eqnarray}
f_q^{[A]} \simeq (z_1)^{(C_A-C_F)/C_A}(y_3)^{C_F/C_A}  \rightarrow 0
\end{eqnarray}
for $z_1\rightarrow 0$.  

For $a=g$, we have
\begin{eqnarray}
L_g^{[A]} \simeq  L_q^{[B1]} \simeq -L_q^{[C1]} \simeq 4C_A^2{K(z_3) \over z_1}. \end{eqnarray}
Therefore, ${\tilde J}^{(g)}_0$ for  $z_1 \ll z_2,z_3$ is approximated as
\begin{eqnarray}
{\tilde J}^{(g)}_0 \simeq 4C_A^2{K(z_3) \over z_1}\log {z_1 \over z_3}
\end{eqnarray}
and 
\begin{eqnarray}
{\tilde J}^{(g)}_0 - \sum_{j=B1,B2}L^{[j]}_g{\rm log}f_g^{[j]} \simeq 4C_A^2{K(z_3) \over z_1}\log y_3,
\end{eqnarray}
which gives 
\begin{eqnarray}
f_g^{[A]} \rightarrow y_3
\end{eqnarray}
for $z_1\rightarrow 0$.  

Above results suggest that the virtual contribution for radiated gluons should be properly taken into account in the two-body branches, particularly for gluons radiated by initial state gluons at the NLL order accuracy.
 Therefore,  the transverse momentum of the space-like partons may be affected by the virtuality $l^2$ of the radiated gluons.

Here, the transverse momentum of an initial state partons for a two-body decay branch 
\begin{eqnarray}
a(k_0) \rightarrow g(l)+a(k)~~(a=q,g)
\end{eqnarray}
are given by \cite{rf:5}
\begin{eqnarray}
{\vec k}_T=z{\vec k}_{0T}+{\vec k}_{TR}
\end{eqnarray}
with
\begin{eqnarray}
|{\vec k}_{TR}|^2=z(1-z) \left[k_0^2+{-k^2 \over z}-{l^2 \over 1-z}\right],
\end{eqnarray}
with $z=(kn)/(k_0n)$. Here, the virtuality of the radiated gluon is restricted by $l^2 \leq f_a^{[A]}(-k^2)$.

\section{TMD parton distributions}

In this section, some numerical results of TMD parton momentum distributions  $x_Fa(x_F,k_T^2,M^2)$ with $k_T^2=|{\vec k}_T|^2$ for $a=q,g$ are presented.

First, we calculate on-shell gluon radiations $(l^2=l_0^2 =0.1{\rm GeV^2})$ with $|{\vec k}_T|^2 \simeq (1-z)(-k^2)$ , where the virtuality $k_0^2$ and the transverse momentum ${\vec k}_{0T}$ of a parent parton in Eqs.$(2\cdot 24)$ and $(2\cdot 25)$ are neglected (Case 1), which may correspond to the calculation in Ref.\citen{rf:1}.  As shown in Appendix A,  a factorization scale $M$ is chosen as $(-k^2_F)z_F^2/(1-z_F) \leq M^2$, where $F$ denotes quantity generated in the last step of parton evolutions. 

In Monte-Carlo calculation, initial state parton evolutions are generated by momentum conserved parton shower model at the NLL order of QCD, which has been implemented in previous works \cite{rf:2}, except the choice of kinematical conditions of the effective two-body decay functions and the factorization scale $M$.

In Figs.3 and 4, the calculated results of Case 1 are shown by squared symbols for $x_F=10^{-1}$ and $x_F=10^{-2}$ at $M^2=10^4{\rm GeV}^2$,  where the parton evolution start from $-k^2_0=Q^2_0=1{\rm GeV^2}$ without intrinsic transverse momentum inside hadrons.\footnote{Input parton distributions implemented in the calculations  are  those given in Ref.\citen{rf:6}} Here, the dashed curves represent the results calculated by formula in Ref.\citen{rf:1}.

\begin{figure}
\centerline{\includegraphics[width=10cm]{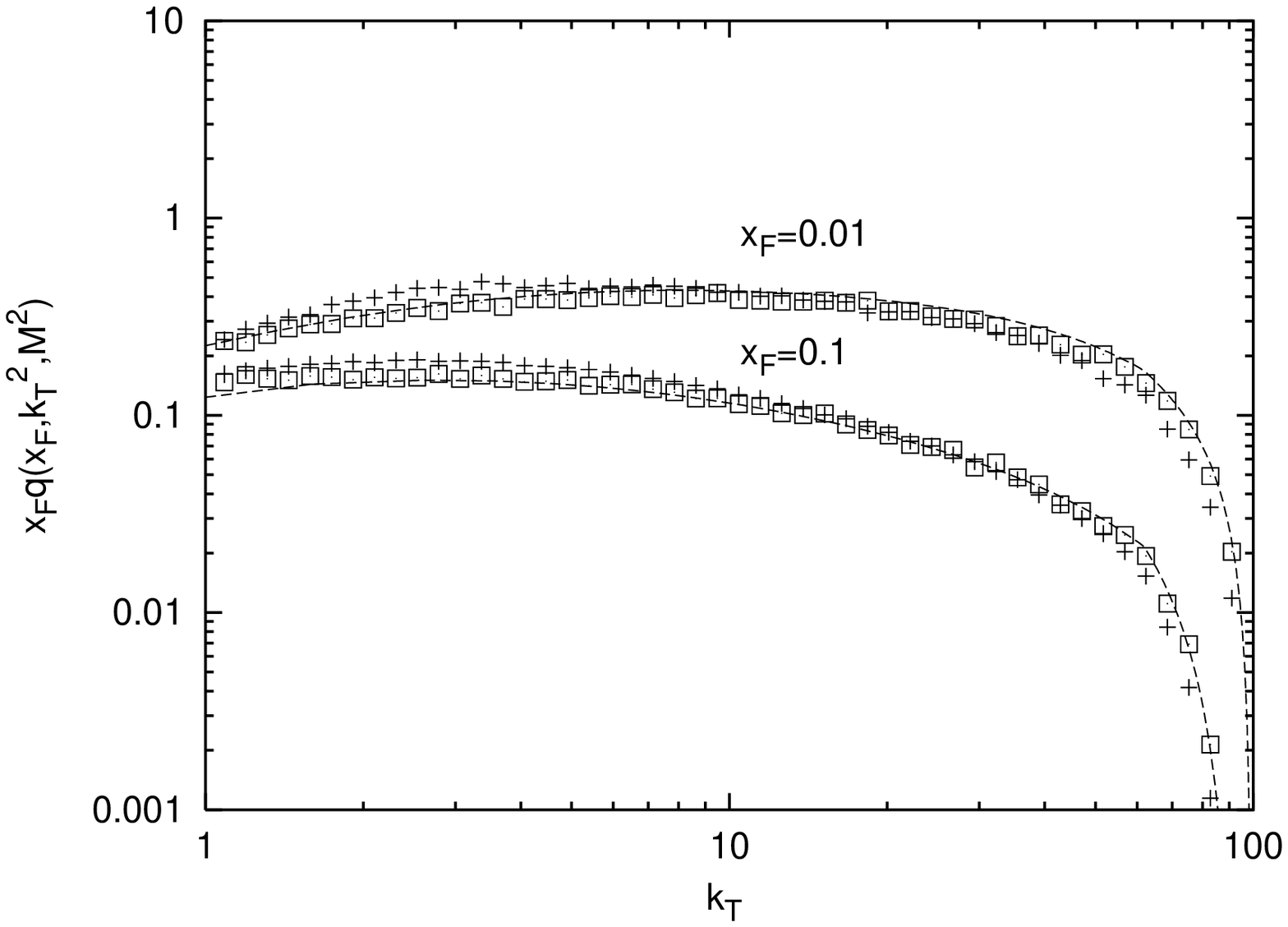}}
\caption{The $k_T$ dependence of the momentum distributions of flavor singlet quarks ($x_Fq$) for $x_F=10^{-1}$ and $10^{-2}$ at $M^2=10^4{\rm GeV}^2$.  The squared symbols and the crossed symbols represent the results for Case 1 and Case 2, respectively. The dashed curves represent the results calculated by formula in Ref.\citen{rf:1}. }
\end{figure}

\begin{figure}
\centerline{\includegraphics[width=10cm]{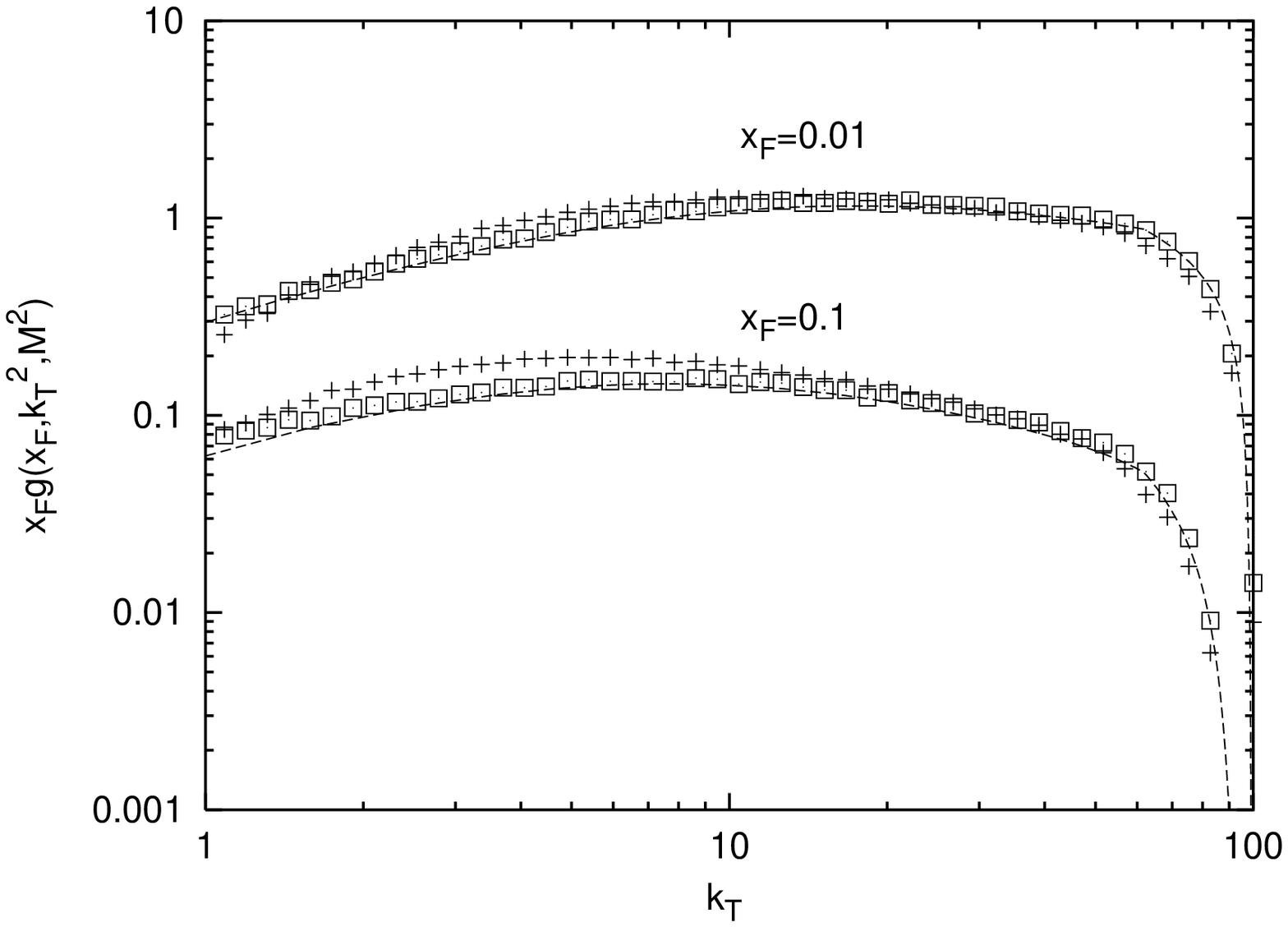}}
\caption{The $k_T$ dependence of the momentum distributions of gluons ($x_Fg$) for $x_F=10^{-1}$ and $10^{-2}$ at $M^2=10^4{\rm GeV}^2$.  The squared symbols and the crossed symbols represent the results for Case 1 and Case 2, respectively. The dashed curves represent the results calculated by formula in Ref.\citen{rf:1}. }
\end{figure}

The results with kinematical conditions (Case 2) in Eqs.$(2 \cdot 11)$ with $(2\cdot 12)$ are represented by plus symbols. 

As shown in the numerical results, even if the angular ordering condition is imposed in initial parton evolutions, the contributions from virtualities of out-going partons , which is a part of the NLL terms, remain for parton radiations.

\section{Summary and comments}
 
   In this paper, we investigated transverse-momentum-dependent (TMD) parton distributions at the next-to-leading logarithmic (NLL) order of QCD based on three-body decay functions for parton branching processes. \cite{rf:3}

In order to generate the initial state parton evolutions, we implemented a parton shower model based on the evolution of the momentum distributions which was extended to the NLL order of QCD. \cite{rf:2}  
   In this type of model, the total momentum of the initial state partons is conserved.   Therefore, it is not necessary to introduce non-trivial weight factors into this model in order to reproduce scaling violation of the flavor singlet parton distributions up to their normalization.   

  In the generation of transverse momenta for the initial state partons with space-like virtuality, the NLL order terms are included in the kinematical conditions for the two-body branching vertices   determined by the three-body decay functions.

As an example, we compared our results with those evaluated at the last step of parton evolutions,\cite{rf:1}, where we implemented the factorization method which is consistent with the angular ordering method in Ref.\citen{rf:1}.  We found that these two methods  give consistent results  for on-shell parton radiations.

 However, it is pointed out that the kinematical boundary of the virtuality $l^2$  for the radiated virtual gluons, which subsequently decays into two gluons, is strongly suppressed by the NLL order contribution for gluons radiated from the initial state quarks.   On the contrary, for gluons radiated by initial gluons,  the kinematical boundary in two-body decay remains as $l^2 \leq (1-z)(-k^2)$ for soft gluon radiations, where $z$ and $k^2$ are a momentum fraction and a vertuality of a space-like parton.

The results presented in this paper suggest that the virtual contribution for radiated gluons are determined by the three-body decay functions and they should be taken into account for gluons radiated from initial state partons at the NLL order accuracy. Therefore, the transverse momentum dependence of parton distributions are affected  by the phase space restriction due to the NLL order contributions in the initial state parton radiations.

  As shown in this paper,  the NLL order terms contribute not only to the evolution of the longitudinal momentum distributions, which is usually taken into account in the evaluation of the scattering cross sections, but also to the transverse momentum distributions for the initial state partons that couple to hard scattering processes.  

The studies presented in this paper may be useful for evaluation of  hadronic cross sections at the NLL order accuracy. More careful studied may be needed for evaluation of the  TMD parton distributions at the NLL order of QCD.

\section*{Acknowledgements}

This work was supported in part by RCMAS (the Research Center for Measurement in Advanced Science) of Rikkyo University.


\vspace{5mm}


\begin{center}
{\Large Appendix A}
\end{center}
\vspace{5mm}

 We consider a sequential radiation of gluons by a virtual parton with the space-like virtuality as
$$ a(p) \rightarrow g(k_1)+a(k) \rightarrow g(k_1)+g(k_2)+a(k_3). $$
for $a=q,g$. Here, we define $k=p-k_1=k_2+k_3, -k^2=-(k_2+k_3)^2=-s_{23},-k_3^2=-s,\eta=1-z_1$ and $\zeta=z_3/\eta$, respectively. Here, $z_i$ is defined in Eq.$(2 \cdot3)$.

The phase space restriction given in Eq.$(2\cdot12)$ as $ -s_{23} < (-s)z_1/(y_3y_1)\simeq (-s)z_1/(z_2y_1) $ for $z_1\ll z_2,z_3$ is written by
\begin{eqnarray*}
-k^2\leq {1-\eta \over \eta^2(1-\zeta)}(-s). 
\end{eqnarray*}
For $-k^2\ll -s$ and $z_1\ll 1$, the transverse momenta of $k$ and $k_3$ may be apploximated  by 
\begin{eqnarray*}
k_T^2 \simeq (1-\eta)(-k^2)  
\end{eqnarray*}
and 
\begin{eqnarray*}
k_{3T}^2 \simeq (1-\zeta)(-s),  
\end{eqnarray*}
respectively. 
Therefore, we have a relation
\begin{eqnarray*}
\eta^2 {k_T^2 \over (1-\eta)^2}\leq {k_{3T}^2 \over (1-\zeta)^2},  
\end{eqnarray*}
which is the same constraint as obtained in Ref.\citen{rf:1}.

If $a(k)\rightarrow g(k_2)+a(k_3)$ branch is the last step of parton evolution, we define the factorization scale $M$ as
$$ \zeta^2{k_{3T}^2 \over (1-\zeta)^2} \simeq  \zeta^2{-k^2 \over 1-\zeta}\leq M^2. $$
The choice of the factorization scale should be compensated by corresponding scale in hard process.


\end{document}